# Morphological analysis of ultra fine $\alpha$-MnO$_2$ nanowires under different reaction conditions

*Niraj Kumar[a], P. Dineshkumar[a], R. Rameshbabu[a], Arijit Sen[ab*]*

[a]*Department of Physics and Nanotechnology, SRM University, Kattankulathur-603203, India*

[b]*SRM Research Institute, SRM University, Kattankulathur-603203, India*

[*]Corresponding author E-mail: arijit.s@res.srmuniv.ac.in (A. Sen).

**Abstract**

A simple hydrothermal method was developed for the synthesis of ultra fine single-crystal $\alpha$-MnO$_2$ nanowires by only using potassium permanganate and sodium nitrite in acidic solution, without any seed or template. Detailed analysis of the obtained nanowires was done using X-ray diffraction (XRD), Fourier transform infrared spectroscopy (FTIR) and high resolution transmission electron microscopy (HRTEM) measurements. The as-prepared $\alpha$-MnO$_2$ nanowires have the average diameter of 10-40nm and a length up to 0.1 to 2µm. Moreover, the effects of reaction temperature, time and reactant molar ratios on the morphology of the final product were studied in detail using field emission scanning electron microscopy (FESEM) measurements.

## 1. Introduction

MnO$_2$ is largely used as cathode material for energy storage such as supercapacitors, high capacity rechargeable lithium batteries for their advantages of low cost, earth abundance, environmental friendliness and superior performance in energy capacity [1, 2]. MnO$_2$ shows high structural diversity and can exist in numerous crystallographic polymorphs, such as $\alpha$-, $\beta$-, $\gamma$ -, $\delta$- and $\varepsilon$-MnO$_2$ [3, 4]. Recent studies have been focused on synthesizing one dimensional (1D) nanostructures and constructing hierarchical structures or networks, which are expected to play an important role in fabricating the next generation of microelectronic and optoelectronic devices since they can function as both building units and interconnections [5]. Therefore the synthesis of pure MnO$_2$ phase with controlled 1D morphologies including nanobelts [6], nanrods [7, 8], nanotubes [9] and nanowires [10-12] have attracted much attention.

Hydrothermal technique is one of the simplest and low cost methods to achieve desired shape and structure of $MnO_2$ [13]. Feng et. al, have reported the one-pot hydrothermal synthesis of $α$-$MnO_2$ nanorods with average diameter of 300 ± 200nm and a length up to 1.2 ± 0.2μm [14]. Ji et. al, have reported the hydrothermal synthesis of $α$-$MnO_2$ nanorods having hollow, polyhedral structure and uniform size with length of about 2-3μm [15]. Kim et. al, have reported the hydrothermal synthesis, structure and magnetic properties of structurally well ordered single crystalline $α$-$MnO_2$ nanorods of 50-100nm diameters and few micro meters in length [16].

Here, a simple one-pot hydrothermal technique is followed to obtain the uniform structure of $α$-$MnO_2$ nanowires. Compared to other soft chemical routes, this technique can precisely control the morphology of the product only by modifying the experimental conditions such as reactant concentration, reaction temperature and time. Its remarkable advantages include simple apparatus, short reaction time, and facile operation conditions with no requirement for any surfactant or template.

## 2. Experimental

### 2.1 Materials

Potassium permanganate ($KMnO_4$), sodium nitrite ($NaNO_2$) and sulfuric acid ($H_2SO_4$) were purchased from Sigma Aldrich with 99.9% purity. Further reagents used were of analytical grade with high purity. De-ionized water was used for preparing all the aqueous solutions.

### 2.2 Preparation of α–$MnO_2$ nanowires

In a typical synthesis, $KMnO_4$ and $NaNO_2$ reagents were mixed in the molar ratio of 2:3 respectively. Then 0.3M $H_2SO_4$ solution was prepared in 3ml of water. The solution was added drop wise in the mixture under continuous stirring to form a homogeneous solution of 37ml. The solution was sealed in teflon-lined stainless steel autoclave (50 ml) of 80% capacity of the total volume. The autoclave was kept in muffle furnace and the hydrothermal process was carried out at a temperature of 180°C for 12h. The final product obtained was washed with de-ionized water for several times and dried in hot air oven. For further purification the product was calcined at 400°C for 4h.

Meanwhile, for comparative study, a series of control experiments were performed by adjusting the reaction parameters such as reaction temperature, time and the molar ratio of permanganate to nitrite ions.

### 2.3 Characterization

The products were characterized by powder X-ray diffraction (XRD) measurement using a PAN analytical X' Pert Pro diffractometer employing Cu-Kα rays of wavelength 1.5406Å with a tube current of 30mA at 40kV in the 2θ range of 10-80 degree. The functional groups of the material was studied using Fourier transform infrared spectroscopy (FTIR) using Perkin Elmer Spectrophotometer by the KBr pellet technique in the range of 400-2000cm$^{-1}$. The morphological analyses of the samples were done by Field emission scanning electron microscopy (Quanta 200 FEG FE-SEM). High resolution Transmission electron microscopy (HR-TEM, JEM-2010, 200kV) was also used to examine crystalline $MnO_2$ nanostructure. The specimen for TEM imaging study was prepared by suspending powder sample in acetone.

## 3. Results and discussions

### 3.1 Structural and Functional group analysis

Fig. 1(a) shows the XRD pattern for the sample prepared at a temperature of 180ºC for 12h with $KMnO_4$ and $NaNO_2$ in the molar ratio of 2:3. The high intensity diffraction peaks at 2θ = 12.7, 18.1, 28.8, 37.4, 49.8, 60.2 can be assigned to a pure tetragonal phase of α-$MnO_2$ (JCPDS 44-0141) and no other characteristic peaks of impurities were observed. This suggests the ultrafine nature of the final product.

Fig. 1(b) shows the FTIR spectrum of the product. The bands at around 1640 and 1387 cm$^{-1}$ correspond to the O-H vibrating mode of traces of absorbed water. The bands at about 718, 527 and 480cm$^{-1}$ that are below 750cm$^{-1}$ can be attributed to the Mn-O vibrations of $MnO_6$ octahedra in α-$MnO_2$. The relatively simple FTIR pattern is ascribed to the highly structural symmetry of α-$MnO_2$. The IR analysis here is found to be in agreement with the results reported in the literature [17, 18].

### 3.2 Microstructure analysis

The TEM and HRTEM images Fig. 2(a-d) revealed that the α-$MnO_2$ obtained at 180ºC exhibits high quality nanowires. The nanowires were well dispersed and maintained one dimensional (an average diameter of 30nm) morphology. The corresponding selected area of the electron diffraction (SAED) pattern Fig. 2(e) supports the formation of high quality tetragonal structure of α-$MnO_2$ nanowires. The high intensity peaks of Manganese and Oxygen elements in Energy dispersive X-ray analysis (EDX) clearly indicates, the obtained nanowires were in pure form with negligible impurities.

### 3.3 Morphological analysis

Fig. 3(a-c) shows the FESEM images at different magnifications of the as prepared nanowires. It further suggests the nanowires had ultra fine shape with negligible agglomerations and having diameters in the range of 10-40nm.

### 3.3.1 Influence of nitrite concentration

Different amounts of sodium nitrite concentration were tested while keeping the other synthesis parameters constant. $MnO_2$ nanowires were formed when $KMnO_4$ and $NaNO_2$ were mixed in the ratio of 2:1 with negligible difference as shown in Fig. 3(d). A noticeable difference was observed on increasing the sodium nitrite concentration. When the molar ratio was changed to 2:5, there was increase in the diameter of nanowires resembling nanorods like structure as observed from Fig. 3(e). This increase in diameter was in the range of 50-100nm. On further increasing the nitrite concentration by changing the molar ratio to 2:7 resulted in the formation of different types of nanostructures like nanograins, nanorods and nanowires simultaneously. This is clearly depicted in Fig. 3(f).

### 3.3.2 Influence of reaction temperature

Effect of varying reaction temperatures of 60ºC, 100ºC and 140ºC were also analyzed while keeping the other parameters constant. FESEM image of the sample synthesized at 60ºC is shown in Fig. 3(g). At this reaction temperature no nanowires were visible. It shows nanograins like structure. Fine nanowires were visible at the reaction temperature of 100ºC but with little agglomerations as shown in Fig. 3(h). On increasing the reaction temperature there was reduction in the agglomerations. At the reaction temperature of 140ºC, nanowires were visible with negligible agglomerations as shown in Fig. 3(i).

### 3.3.3 Influence of reaction time

Fig. 3(j-l) shows the FESEM images of the samples prepared at different reaction times of 3, 6 and 9h respectively, with no change in other reaction parameters. It can be clearly observed that with the increasing reaction time agglomerations formed in the nanowires were greatly reduced. Highly agglomerated nanowires were visible at the reaction time of 3h. Small and large reductions in agglomerations of nanowires were observed at the reaction time of 6 and 9h respectively. The effect of reaction time was found to be similar with the effect of reaction

temperature. Hence the varying temperatures and times both had an impact on the extent of agglomerations in the nanowires, whereas the varying nitrite concentration resulted in the change of shapes of final products.

**4. Conclusion**

Ultrafine $α$-$MnO_2$ nanowires were synthesized by a facile hydrothermal method without using any templates, seeds or other capping agents. The influence of changing the reaction time and temperature, the molar ratio of permanganate to nitrite were studied systematically. A high yield and ultrafine nanowires were obtained under optimized conditions with 2:3 molar ratio of permanganate to nitrite in acid media at 180$^o$C for 12h. The varying reaction time, temperature and nitrite concentration lead to the morphological changes in the final product.

**References**

[1] Tang W, Hou YY, Wang XJ, Bai Y, Zhu YS, Sun H, Yue YB, Wu YP, Zhu K, Holze R. J Power Sources 2012; 197: 330-3.

[2] Yang Y, Xiao L, Zhao Y, Wang F. Int J Electrochem Sci 2008; 3: 67-74.

[3] Gao T, Fjellvag H, Norby P. Nanotec 2009; 20: 055610 (7pp).

[4] Post JE. Proc Natl Acad Sci USA 1999; 96: 3447-54.

[5] Cheng F, Zhao J, Song W, Li C, Ma H, Chen J, Shen P. Inorg Chem 2006; 45: 2038-44.

[6] Li G, Liang L, Pang H, Peng H. Mater Lett 2007; 61: 3319-22 .

[7] Yang R, Wang Z, Dai L, Chen L. Mater Chem Phys 2005; 93: 149-53.

[8] Tang N, Tian X, Yang C, Pi Z, Han Q. J Phys Chem Solids 2010; 71: 258-62.

[9] Luo J, Zhu HT, Fan HM, Liang JK, Shi HL, Rao GH, Li JB, Du ZM, Shen ZK. J Phys Chem C 2008; 112: 12594-8.

[10] Zhang LC, Liu ZH, Lv H, Tang X, Ooi K. J Phys Chem C 2007; 111: 8418-23.

[11] Wang X, Li Y. J Am Chem Soc 2002; 124: 2880-81.

[12] Li Q, Olson JB, Penner RM. Chem Mater 2004; 16: 3402-5.

[13] Wu J, Huang H, Yu L, Hu J. Adv Mater Phy Chem. 2013; 3: 201-5.

[14] Feng J, Zhang P, Wang A, Zhang Y, Dong W, Chen J. J Coll Int Sci 2011; 359: 1-8.

[15] Ji Z, Dong B, Guo H, Chai Y, Li Y, Liu Y, Liu C. Mater Chem Phys 2012; 136: 831-6.

[16] Kim H, Lee J, Kim Y, Jung M, Jaglicic Z, Umek P, Dolinsek J. Nano Res Lett 2007; 2: 81-6

[17] Yang R, Wang Z, Dai L, Chen L. Mater Chem Phys 2005; 93: 149-153.

[18] Wang H, Lu Z, Qian D, Li Y, Zhang W. Nanotec 2007; 18: 115616 (5pp).

**Figure Captions**

Fig. 1. (a) XRD pattern and (b) FTIR spectroscopy of the final product prepared at a temperature of 180$^{o}$C for 12h with KMnO$_4$ and NaNO$_2$ in the molar ratio of 2:3.

Fig. 2. (a-d) HRTEM images at different magnifications, (e) SAED pattern and (f) EDAX analysis of the final product prepared at a temperature of 180$^{o}$C for 12h with KMnO$_4$ and NaNO$_2$ in the molar ratio of 2:3.

Fig. 3. FESEM images of the final product prepared at a temperature of 180$^{o}$C for 12h with KMnO$_4$ and NaNO$_2$ in the molar ratio of 2:3, (a), (b) and (c) at different magnifications; the products obtained at different molar ratios of (d) 2:1, (e) 2:5 and (f) 2:7 at a temperature of 180$^{o}$C for 12h; the products obtained at varying temperatures of (g) 60$^{o}$C, (h) 100$^{o}$C, and (i) 140$^{o}$C in the molar ratio of 2:3 for 12h and the products obtained for times (j) 3h, (k) 6h and (l) 9h in the molar ratio of 2:3 at a temperature of 180$^{o}$C.

**Figures**

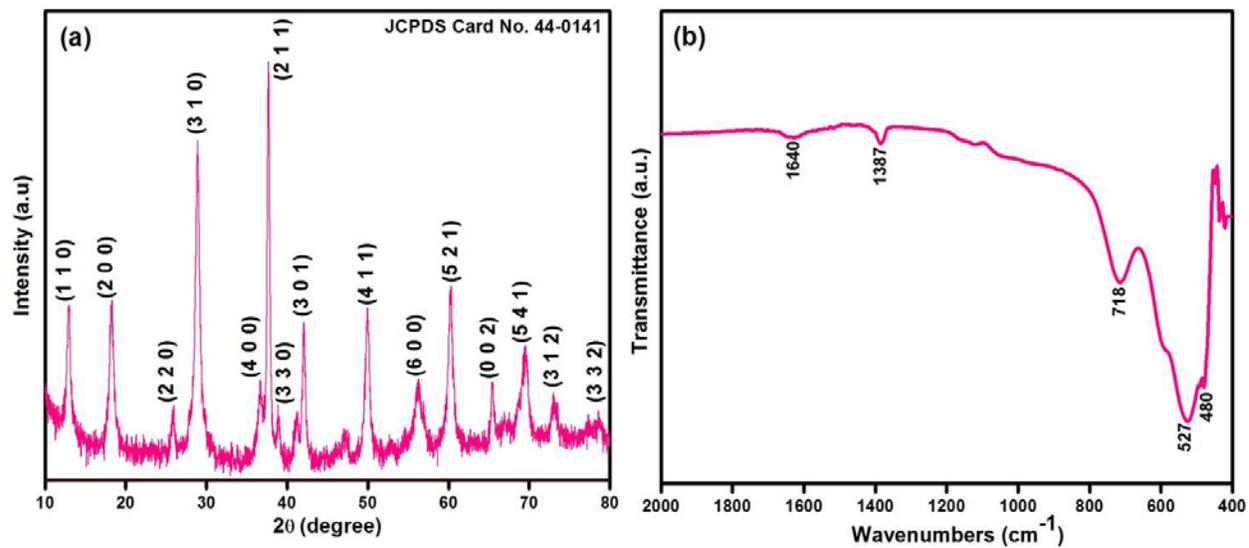

**Figure 1**

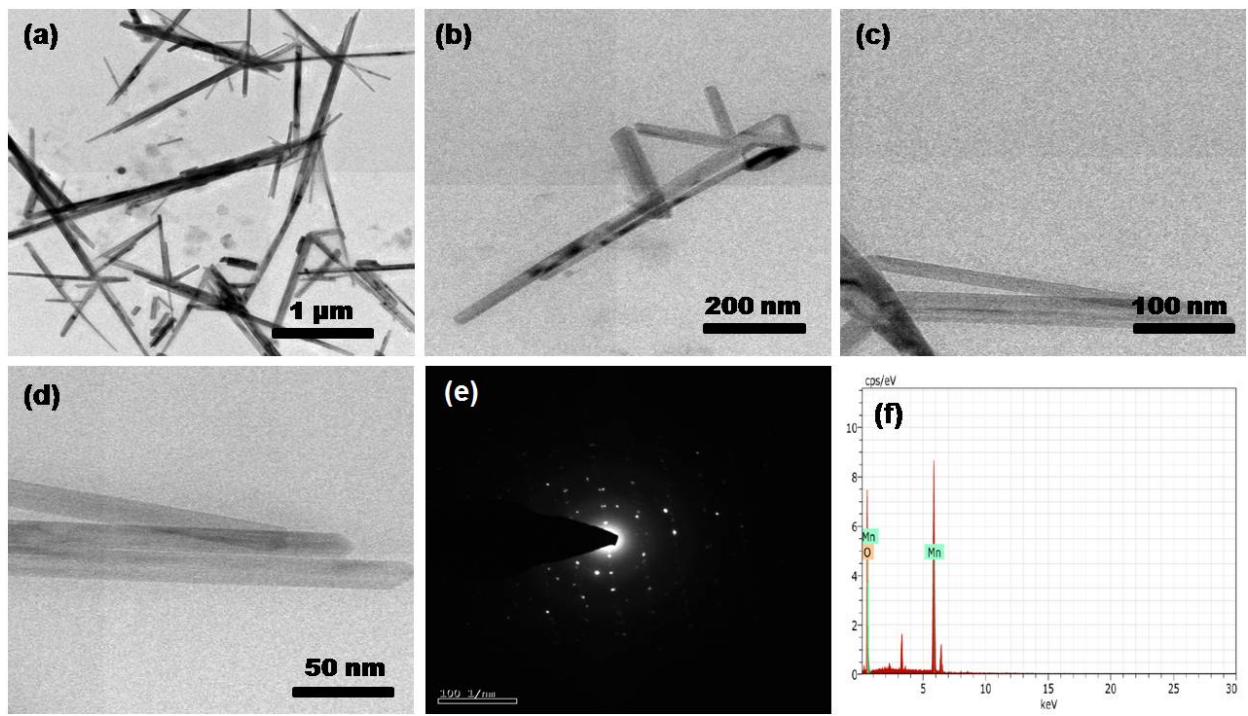

**Figure 2**

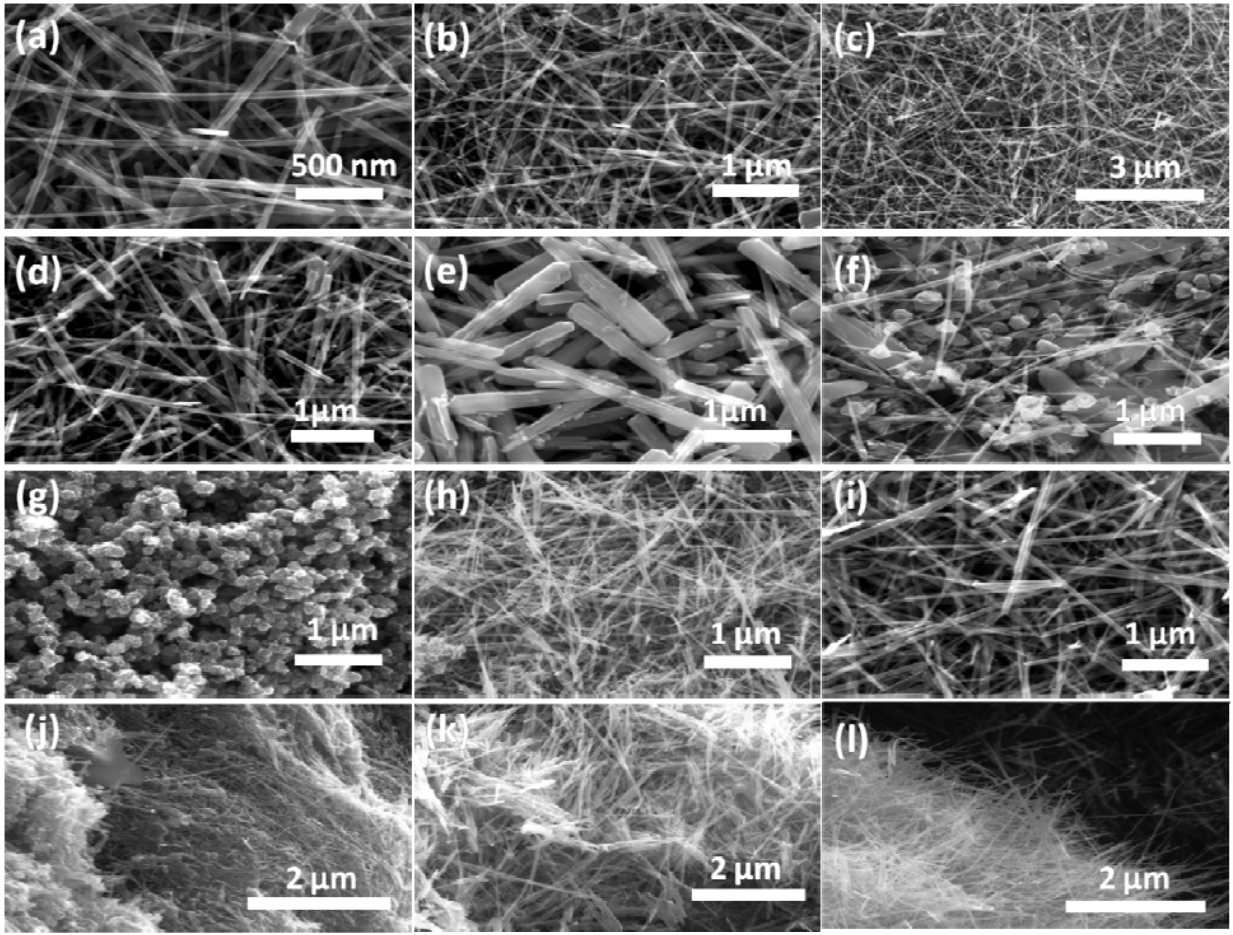

**Figure 3**